\documentstyle[12pt]{article}
\title{Introduction to the Effective Field Theory Description of Gravity}
\author{John F. Donoghue\\ [5mm]
Department of Physics and Astronomy\\
University of Massachusetts, Amherst, MA ~01003}

\date{ }
\begin{document}
\begin{titlepage}
\maketitle
\begin{abstract}
This is a pedagogical introduction to the treatment of general relativity as a
quantum effective field theory.  Gravity fits nicely into the effective field
theory description and forms a good quantum theory at ordinary energies.
\end{abstract}

{\vfill Lectures presented at the Advanced School on Effective Theories,
Almunecar, Spain June 1995 (to be published in the proceedings).
UMHEP-424, gr-qc/9512024}

\end{titlepage}

\section{Introduction}
  The conventional wisdom is that general relativity and quantum
mechanics
are presently incompatible. Of the ``four fundamental forces''
gravity is said to be different because a
quantum version of the theory does not exist. We feel less
satisfied with the theory of gravity and exclude it from being
recognized as a full member of
the Standard Model. Part of the trouble is that
we have tried to unnaturally force gravity into the mold of
renormalizable field theories. In the old way of thinking,
only the class of renormalizable field theories were considered
workable quantum theories. For this reason, general relativity
was considered a failure as a quantum field theory. However
we now think differently about renormalizability. So-called
non-renormalizable theories can be renormalized if treated in
a general enough framework, and they are not inconsistent with
quantum mechanics[1]. In the framework of effective field theories[2],
the effects of quantum physics can be analyzed and reliable
predictions can be made. We will see that in this regard the
conventional wisdom about gravity is not correct; quantum predictions
can be made.

The key point of effective field theory is the separation of
known physics at the scale that we are working from unknown
physics at much higher energies. Experience has shown that as
we go to higher energies, new degrees of freedom and new
interactions come into play. We have no reason to suspect that
the effects of  our present theory are the whole story at the
highest energies. Effective field theory allows us to
make predictions at present
energies without making unwarranted assumptions about what is
going on at high energies. In addition, whatever the physics of high
energy really is, it will leave residual effects at low energy
in the form of highly suppressed non-renormalizable interactions.
These can be treated without disrupting the low energy theory.
 The use of effective field theory is
not limited to non-renormalizable theories. Even renormalizable
theories benefit from this paradigm. For example, there are the
well-known divergences in all field theories. If these divergences
were really and truly infinite, the manipulations
that we do with them would be nonsense. However we do not believe
that our calculations of these divergences are really correct,
and if our theory is only a low energy effective field theory
of the ultimate finite theory of everything, the manipulations
are perfectly reasonable with the end predictions being
independent of the physics at very high scales. Our feeling
of the reality of the radiative corrections has also convinced
many that it is most natural if the present Standard Model
is an effective theory which breaks down at the TeV scale,
where Higgs self-interactions would otherwise become unnaturally
large. We have even found it useful in Heavy Quark Effective Theory
to convert a renormalizable theory into a non-renormalizable one
in order to more efficiently display the relevant degrees of
freedom and interactions.

 In the case of gravity, we feel that the low energy degrees
of freedom and interactions are those of general relativity.
It would be a surprise if these could not be treated quantum mechanically.
To be sure, radiative corrections appear to involve all energies,
but this is a problem that the effective field theory formalism
handles automatically. We will see that gravity very naturally
fits into the framework of effective field theory[3]. In fact
it is potentially even a {\it better} effective theory than the
Standard Model as the quantum corrections are very small and
the theory shows no hint of a breakdown before the Planck scale.
If we insist on treating general relativity as the isolated
fundamental theory even at very high energies, there will be
the usual problems at high energy. However, the main point is
that we can use the degrees of freedom that we have at ordinary
energies to make quantum calculations relevant for those scales.

In these lecture notes, I will briefly review the structure
of general relativity and its status as a classical effective
field theory. Then I discuss the quantization of the theory,
following the work of
't Hooft and Veltman[4]. The methodology of effective
field theory is explained in regards to the renormalization
of the theory and the extraction of low energy quantum
predictions. I describe the example of the gravitational
interaction of two heavy masses to illustrate the method[3,5].
Some unique features of the gravitational effective theory are
briefly mentioned.

\section{Basic Structure of General Relativity}

Since this is a school primarily for particle physicists, I need to
briefly review general relativity[6,7]. This also allows us
to specify our notation and make a few preliminary comments which are
relevant for the effective field theory treatment.
When a field theorist describes the ingredients of general
relativity, it is
interesting to see how much the description differs from
that of conventional
relativists.  Below is a minimalist field theoretic
presentation of the structure
of the theory.

Lorentz invariance is a global coordinate change which leaves the
Minkowski metric tensor invariant.

\begin{eqnarray}
x'^{\mu} & = & \Lambda^{\mu}_{~\nu} x^{\nu} \nonumber \\
\eta_{\mu \nu} & = & \Lambda^{\alpha}_{~\mu} \Lambda^{\beta}_{~\nu}
\eta_{\alpha \beta} \nonumber \\
\nonumber \\
\eta_{\mu \nu} & = & diag (1, -1, -1, -1) \nonumber \\
d \tau^2 & = & \eta_{\mu \nu} dx^{\prime \mu} dx^{\prime \nu} =
\eta_{\alpha \beta} dx^{\alpha} dx^{\beta}
\end{eqnarray}

\noindent Fields transform as scalars, vectors, etc., under this change

\begin{eqnarray}
\phi^{\prime} (x^{\prime}) & = & \phi(x) \nonumber \\
A^{\prime \mu} (x^{\prime}) & = & \Lambda^{\mu}_{~\nu} A^{\nu} (x)
\end{eqnarray}

\noindent This can be made into a local coordinate change

\begin{eqnarray}
x^{\prime \mu} & = & x^{\prime \mu} (x) \nonumber \\
dx^{\prime \mu} & = & \Lambda^{\mu}_{~\nu} (x) dx^{\nu} \nonumber
\\
\bar{\Lambda}_{\mu} ^{~\nu} (x) & \equiv & \left[ \Lambda^{\mu}
_{~\nu}
(x) \right]^{-1} \nonumber \\
\Lambda^{\mu} _{~\nu} \bar{\Lambda}_{\rho} ^{~\nu} & = &
\delta^{\mu}
_{\rho}
\end{eqnarray}

\noindent but only if the metric is allowed to be a coordinate dependent field
transforming as

\begin{eqnarray}
g^{\prime}_{\mu \nu} (x^{\prime}) & = & \bar{\Lambda}_{\mu}
^{~\alpha}
\bar{\Lambda}_{\nu} ^{~\beta} g_{\alpha \beta} (x) \nonumber \\
d \tau^2 & = & g^{\prime}_{\mu \nu} (x^{\prime}) dx^{\prime \mu}
dx^{\prime \nu} = g_{\alpha \beta} (x) dx^{\alpha} dx^{\beta}
\end{eqnarray}

\noindent with inverse $g^{\mu \nu}$

\begin{equation}
g^{\mu \nu}{(x)} g_{\nu \rho} (x) = \delta^{\mu} _{\rho}
\end{equation}

\noindent Scalar and vector fields are now defined with the properties

\begin{eqnarray}
\phi^{\prime} (x^{\prime}) & = & \phi (x) \nonumber \\
A^{\prime \mu} (x^{\prime}) & = & \Lambda^{\mu} _{~\nu} (x) A^{\nu}
(x)
\end{eqnarray}

\noindent A covariant derivative can be defined with the right transformation
property (i.e., $D^{\prime}_{\mu} A^{\lambda} = \bar{\Lambda}_{\mu}
^{~\nu} \Lambda^{\lambda} _{~\sigma} D_{\nu} A^{\sigma}$) by

\begin{equation}
D_{\mu} A^{\lambda} = \partial_{\mu} A^{\lambda} +
\Gamma^{\lambda}_{\mu \nu} A^{\nu}
\end{equation}

\noindent where the connection $\Gamma^{\lambda}_{\mu \nu}$ is defined
as

\begin{equation}
\Gamma^{\lambda}_{\mu \nu} = {1 \over 2} g^{\lambda \sigma} \left[
\partial_{\mu} g_{\sigma \nu} + \partial_{\nu} g_{\sigma \mu} -
\partial_{\sigma} g_{\mu \nu} \right]
\end{equation}

\noindent It is important for the
effective theory that the connection involves
one derivative of the metric $(\Gamma \sim  \partial g)$.

 Similarly we can
define tensor and scalar fields, the curvatures, which depend only on
two derivatives of the metric

\begin{eqnarray}
[D_{\mu}, D_{\nu}] A_{\alpha} & \equiv & R^{\beta} _{~\alpha \mu \nu}
A_{\beta} \nonumber \\
\nonumber \\
R^{\beta} _{~\alpha \mu \nu} & = & \partial_{\mu} \Gamma^{\beta}
_{~\alpha \nu} - \partial_{\nu} \Gamma^{\beta} _{~\alpha \mu} +
\Gamma^{\lambda}  _{~\alpha \nu} \Gamma^{\beta} _{~\lambda \mu} -
\Gamma^{\lambda} _{~\alpha  \mu} \Gamma^{\beta} _{~\lambda \nu}
\nonumber \\
R _{\alpha \mu} & \equiv & R^{\lambda} _{~\alpha \mu \lambda}
\nonumber \\
R & \equiv & g^{\alpha \mu} R_{\alpha \mu}
\end{eqnarray}

\noindent The curvature is nonlinear in the field $g_{\mu \nu}$.  Despite the
similarity to the construction of the field strength tensor of
Yang Mills field
theory, there is the important difference that the curvatures involve two
derivatives of the basic field $(R \sim \partial \partial g)$.

It is easy to construct an action for the matter fields which is
invariant under
the general coordinate transformation, simply by modifying the usual
Lagrangian to use covariant derivatives and to raise and lower Lorentz
indices with $g_{\mu \nu} (x)$.  In addition, if we want the metric to be a
dynamical field we need an action involving derivatives on the metric.  The
simplest invariant function is the scalar curvature so that one would
postulate

\begin{equation}
S_g = \int d^4 x \sqrt{g} {2 \over \kappa^2} R
\end{equation}

\noindent with $\kappa^2$ presently an unknown constant.  We will return to
this step in the next
section.  Variation of the full action leads to Einstein's
Equation

\begin{eqnarray}
R_{\mu \nu} - {1 \over 2} g_{\mu \nu} R = - 8 \pi G T_{\mu \nu}
\nonumber \\
\sqrt{g} T^{\mu \nu} \equiv - 2 {\partial \over \partial g_{\mu \nu} }
\left( \sqrt{g} {\cal L}_m \right)
\end{eqnarray}

\noindent where ${\cal L}_m$ is the Lagrange density for matter, and
$T^{\mu \nu}$ is the corresponding energy momentum tensor, and $\kappa^2
\equiv 32 \pi G$.  By investigating solutions to this equation, it can be seen
to describe Newtonian gravity in the appropriate limit if $G$ is identified as
the Cavendish constant.

In this summary, invariance requirements take precedence over geometrical
ideas and indeed the fact that this is a good theory for gravity appears only
at the end of this construction.

We will use a few other facts of general relativity which deserves to be
mentioned in this section.  In the weak field limit we can expand the metric
around Minkowski space introducing the dynamical part of the metric as
$h_{\mu \nu}$

\begin{eqnarray}
g_{\mu \nu} & \equiv & \eta_{\mu \nu} + \kappa h_{\mu \nu} \, \qquad
(exactly)
\nonumber \\
g^{\mu \nu} &=& \eta^{\mu \nu} - \kappa h^{\mu \nu} + \kappa^2 h^{\mu
\lambda} h_{\lambda}^{~\nu} + \ldots
\end{eqnarray}

\noindent The weak field gauge invariance is given by

\begin{eqnarray}
x^{\prime \mu} &=& x^{\mu} + \epsilon^{\mu} (x) \, \qquad \epsilon <<
1
\nonumber \\
h_{\mu \nu}^{~\prime} (x^{\prime}) &=& h_{\mu \nu} (x) -
\partial_{\mu}
\epsilon_{\nu} - \partial_{\nu} \epsilon_{\mu}
\end{eqnarray}

\noindent and the curvatures are

\begin{eqnarray}
R_{\mu \nu} &=& {\kappa \over 2} \left[ \partial_{\mu} \partial_{\nu}
h^{\lambda} _{~\lambda} + \partial_{\lambda} \partial^{\lambda} h_{\mu
\nu} - \partial_{\mu} \partial_{\lambda} h^{\lambda} _{~\nu} -
\partial_{\lambda} \partial_{\nu} h^{\lambda} _{~\mu} \right] + {\cal O}
(h^2) \nonumber \\
R &=& \kappa \left[ \Box h^{\lambda} _{~\lambda} - \partial_{\mu}
\partial_{\nu} h^{\mu \nu} \right] + {\cal O} (h^2)
\end{eqnarray}

\noindent where indices are raised and lowered with $\eta_{\mu \nu}$.
This can equally well be done around any fixed smooth
background space time
metric.

 The Greens function does not exist without a gauge choice and it is
most convenient to use harmonic gauge

\begin{equation}
\partial^{\lambda} h_{\mu \lambda} = {1 \over 2} \partial_{\mu}
h^{\lambda} _{~\lambda}
\end{equation}

\noindent which reduces Einstein's Equation in the weak field limit to

\begin{equation}
\Box h_{\mu \nu} = -16 \pi G \left( T_{\mu \nu} - {1 \over 2} \eta_{\mu
\nu} T^{\lambda} _{~\lambda} \right)
\end{equation}

\noindent This has the solution for a static point mass of

\begin{equation}
h_{\mu \nu} = diag (1, 1, 1, 1) \left( 2 {G M \over r} \right)
\end{equation}

\noindent There are also plane wave solutions.  These satisfy

\begin{equation}
R_{\mu \nu} = 0 = \Box h_{\mu \nu}
\end{equation}

\noindent resulting in

\begin{equation}
h_{\mu \nu} = N \epsilon_{\mu \nu} e^{-i p \cdot x} + h. c.
\end{equation}

\noindent with $p^2 = 0$.  The harmonic gauge choice plus residual
gauge freedom can reduce the polarization vector to two transverse traceless
degrees of freedom appropriate for a massless spin two degrees of freedom.

The gravity waves also carry energy and momentum; hence gravity is
nonlinear.  The energy momentum tensor of the gravity waves simplifies a
bit in harmonic gauge and can be put in the form

\begin{eqnarray}
T_{\mu \nu} &=& -{1 \over 4} h_{\alpha \beta} \partial_{\mu}
\partial_{\nu} h^{\alpha \beta} + {1 \over 8} h \partial_{\mu} \partial_{\nu}
h \nonumber \\
& & + {1 \over 8} \eta_{\mu \nu} \left( h^{\alpha \beta} \Box h_{\alpha
\beta} - {1 \over 2} h \Box h \right) \nonumber \\
& & - {1 \over 4} \left( h_{\mu \rho} \Box h^{\rho} _{~\nu} + h_{\nu
\rho}  \Box h^{\rho} _{~\mu} - h_{\mu \nu} \Box h \right) \nonumber \\
& & + {1 \over 8} \partial_{\mu} \partial_{\nu} \left\{ h_{\alpha \beta}
h^{\alpha \beta} - {1 \over 2} h h \right\} - {1 \over 16} \eta_{\mu \nu}
\Box \left\{ h_{\alpha \beta} h^{\alpha \beta} - {1 \over 2} hh \right\}
\nonumber \\
& & - {1 \over 4} \partial_{\alpha} \left[ \partial_{\nu} \left\{ h_{\mu
\beta}  h^{\alpha \beta} \right\} + \partial_{\mu} \left\{ h_{\nu \beta}
h^{\alpha \beta} \right\} \right] \nonumber \\
& & + {1 \over 2} \partial_{\alpha} \left[ h^{\alpha \beta} \left(
\partial_{\nu} h_{\mu \beta} + \partial_{\mu} h_{\nu \beta} \right) \right]
\end{eqnarray}

\noindent with $h \equiv h^{\lambda} _{~\lambda}$.  In this form only the
first term contributes to the forward matrix element of a physical transverse
traceless mode.

\section{Classical Effective Field Theory}

Let us revisit a crucial  step in the derivation of general
relativity.  What is
the rationale for choosing the gravitational action proportional to $R$ and
only $R$?  It is not due to any symmetry and, unlike other theories, cannot
be argued on the basis of renormalizability.  However physically the
curvature is small so that in most applications $R^2$ terms would be yet
smaller.  This leads to a rationale based on classical effective field theory.

There are in fact infinitely many terms allowed by general coordinate
invariance, i.e.,

\begin{equation}
S = \int d^4 x \sqrt{g} \left\{ \Lambda + {2 \over \kappa^2} R + c_1 R^2 + c_2
R_{\mu \nu} R^{\mu \nu} + \ldots + {\cal L}_{matter} \right\}
\end{equation}

\noindent Here the gravitational Lagrangians have been ordered in a
derivative expansion with $\Lambda$ being of order $\partial^0, R$ of
order $\partial^2, R^2$ and $R_{\mu \nu} R^{\mu \nu}$ of order
$\partial^4$ etc.  Note that in four dimensions we do not need to include a
term $R_{\mu \nu \alpha \beta} R^{\mu \nu \alpha \beta}$ as the Gauss
Bonnet theorem allows this contribution to the action to be written in terms
of $R^2$ and $R_{\mu \nu} R^{\mu \nu}$.

The first term in Eq.21 , i.e., $\Lambda$, is related to the cosmological
constant, $\lambda = -8 \pi  G \Lambda$, with Einstein's equation
becoming

\begin{equation}
R_{\mu \nu} - {1 \over 2} g_{\mu \nu} R = -8 \pi  G T_{\mu \nu} -
\lambda g_{\mu \nu}
\end{equation}

\noindent This is a term which in principle should be included, but
cosmology bounds $\mid \lambda \mid < 10^{-56} cm^{-2}, \mid \Lambda
\mid < 10^{-46} GeV^4$ so that this constant is unimportant at ordinary
energies[8].  We then set $\Lambda = 0$ from now on.

In contrast, the $R^2$ terms are able to be shown to be unimportant in a
natural way.  Let us drop Lorentz indices in order to focus on the important
elements, which are the numbers of derivatives.  A $R + R^2$ Lagrangian

\begin{equation}
{\cal L} = {2 \over \kappa^2} R + cR^2
\end{equation}

\noindent has an equation of motion which is of the form

\begin{equation}
\Box h + \kappa^2 c^2 \Box \Box h = 8 \pi GT
\end{equation}

\noindent The Greens function for this wave equation has the form

\begin{eqnarray}
G(x) & = & \int {d^4q \over (2 \pi)^4 } {e^{iq \cdot x} \over q^2
+ \kappa^2 c
q^4} \nonumber \\
& = & \int {d^4 q \over (2 \pi )^4} \left[ {1 \over q^2} - {1 \over q^2 +
{1/\kappa^2 c}} \right] e^{-iq \cdot x}
\end{eqnarray}

\noindent The second term appears like a massive scalar, but with the wrong
overall sign, and leads to a short-ranged Yukawa potential

\begin{equation}
V(r) = - G m_1 m_2 \left[ {1 \over r} - {e^{-r/ \sqrt{\kappa^2 c}} \over
r}\right] .
\end{equation}

\noindent The exact form has been worked out by Stelle[9], who gives the
experimental bounds $c_1, c_2 < 10^{74}$.  Hence, if $c_i$ were a
reasonable number there would be no effect on any observable physics.
[Note that if $c \sim 1, \sqrt{\kappa^2 c} \sim 10^{-35}m]$.  Basically the
curvature is so small that $R^2$ terms are irrelevant at ordinary scales.

As a slightly technical aside, in an effective field theory we should
not treat the $R^2$ terms to all orders,
as is done above in the exponential of the Yukawa solution,
but only include the
first corrections in $\kappa^2c$.  This is because at higher orders
in $\kappa^2c$ we
would also be sensitive to yet higher terms in the effective Lagrangian
($R^3, R^4$ etc.) so that we really do not know the full $r \rightarrow 0$
behavior.  Rather, for $\sqrt{\kappa^2c}$ small we can note the Yukawa
potential becomes a representation of a delta function

\begin{equation}
{e^{-r/ \sqrt{\kappa^2c}} \over r} \rightarrow 4\pi
\kappa^2c \delta^3 (\vec{r})
\end{equation}

\noindent Alternatively in the Greens function we could note that

\begin{equation}
{1 \over q^2 + \kappa^2c q^4} = {1 \over q^2} - \kappa^2c + \cdots
\end{equation}

\noindent and that the Fourier transform of a constant is a delta function.
Either way, one is lead to a form of the potential

\begin{equation}
V(r) = -Gm_1 M_2 \left[{1 \over r} + 128 \pi^2 G (c_1 - c_2) \delta^3
(\vec{x}) \right]
\end{equation}

\noindent $R^2$ terms in the Lagrangian lead to a very weak and short
range modification to the gravitational interaction.

Thus when treated as a classical effective field theory, we can start with the
more general Lagrangian, and find that only the effect of the Einstein action,
$R$, is visible in any test of general relativity.  We need not make any
unnatural restrictions on the Lagrangian to exclude $R^2$ and $R_{\mu
\nu} R^{\mu \nu}$ terms. J. Simon[10] has shown that the standard problems
with classical $R + R^2$ gravity are not problems when one restricts
onself to the low energy domain appropriate for an effective field
theory.

\section{Quantization}

There is a beautiful and simple formalism for the quantization of gravity.
The most attractive variant combines the covariant quantization pioneered by
Feynman and De Witt[11] with the
background field method[12] introduced in this
context by 't Hooft and Veltman[4].  The quantization of a gauge theory
always involves fixing a gauge. This can in principle cause trouble if
this procedure then induces divergences which can not be absorbed
in the coefficients of the most general Lagrangian which displays the
gauge symmetry. The background field method solves this problem
because the calculation retains the symmetry under transformations of
the background field and therefor the loop expansion will be
gauge invariant, retaining the symmetries of general relativity.

Consider the expansion of the metric about a smooth background field
$\bar{g}_{\mu \nu} (x)$,

\begin{equation}
g_{\mu \nu} (x) = \bar{g}_{\mu \nu} (x) + \kappa h_{\mu \nu}
\end{equation}

\noindent Indices are now raised and lowered with $\bar{g}$.  The
Lagrangian may be expanded in the quantum field $h_{\mu \nu}$[4,13].

\begin{eqnarray}
{2 \over \kappa^2} \sqrt{g} R & = & \sqrt{\bar{g}} \left\{ {2
\over \kappa^2}
\bar{R} + {\cal L}_g^{(1)} + {\cal L}_g^{(2)} + \cdots \right\}\nonumber
\\
{\cal L}^{(1)}_g & = & {h_{\mu \nu} \over \kappa} \left[ \bar{g}^{\mu \nu}
\bar{R} - 2 \bar{R}^{\mu \nu} \right] \nonumber \\
{\cal L}^{(2)}_g & = & {1 \over 2} D_{\alpha} h_{\mu \nu} D^{\alpha}
h^{\mu \nu} - {1 \over 2} D_{\alpha} h D^{\alpha} h + D_{\alpha} h
D_{\beta} h^{\alpha \beta} \\
& & -D_{\alpha} h_{\mu \beta} D^{\beta} h^{\mu \alpha} + \bar{R} \left(
{1 \over 2} h^2 - {1 \over 2} h_{\mu \nu}h^{\mu \nu} \right) \nonumber \\
& & + \bar{R}^{\mu \nu} \left(2 h^{\lambda} _{~\mu} h_{\nu \alpha} - h
h_{\mu \nu} \right) \nonumber
\end{eqnarray}

\noindent Here $D_{\alpha}$ is a covariant derivative with respect to the
background field.  The total set of terms linear in $h_{\mu \nu}$ (including
those from the matter Lagrangian) will vanish if $\bar{g}_{\mu \nu}$
satisfies Einstein's equation.  We are then left with a quadratic Lagrangian
plus interaction terms of higher order.

However, the quadratic Lagrangian cannot be quantized without gauge
fixing and the associated Feynman-DeWitt-Fadeev-Popov
ghost fields[11,14].  Let
us briefly recall the logic, which is the same as for the quantization of Yang
Mills theories[14,15].  The path integral overcounts fields which are
equivalent under a gauge
transformation and we need to divide out extra gauge copies.  Carrying out
the procedure leads to a path integral

\begin{equation}
Z = \int d h_{\mu \nu} \delta (G_{\alpha} (h)) det \mid {\partial G_{\alpha}
\over \partial \epsilon_{\beta}} \mid e^{iS}
\end{equation}

\noindent where $G_{\alpha} (h)$ is the gauge constraint to be imposed and
${\partial G_{\alpha} \over \partial \epsilon_{\beta}}$ refers
to the variation
of the constraint under infinitesimal gauge transformation as in Eq.13 .
Exponentiation of $\delta (G_{\mu} (h))$ leads to the addition of a gauge
fixing term, ${\cal L}_{gf}$, to the quadratic Lagrangian.  Exponentiation
of $det \mid {\partial G \over \partial \epsilon} \mid$ by introducing new
fermion fields and using

\begin{equation}
det M = \int d \eta d \bar{\eta} e^{i \int d^4x \bar{\eta} M \eta}
\end{equation}

\noindent brings in the ghost Lagrangian and completes the procedure of
producing a quadratic Lagrangian with gauge fixing.

In this case, we would like to impose the harmonic gauge constraint in the
background field, and can choose the constraint[4]

\begin{equation}
G^{\alpha} = \sqrt[4]{g} \left( D^{\nu} h_{\mu \nu} - {1 \over 2} D_{\mu}
h^{\lambda} _{~\lambda} \right) t^{\nu \alpha}
\end{equation}

\noindent where

\begin{equation}
\eta_{\alpha \beta} t^{\mu \alpha} t^{\nu \beta} = \bar{g}^{\mu \nu}
\end{equation}

\noindent This leads to the gauge fixing Lagrangian[4]

\begin{equation}
{\cal L}_{gf} = \sqrt{ \bar{g}} \left\{ \left( D^{\nu} h_{\mu \nu} - {1 \over
2} D_{\mu} h^{\lambda} _{~\lambda} \right) \left( D_{\sigma} h^{\mu
\sigma} - {1 \over 2} D^{\mu} h^{\sigma} _{~\sigma} \right) \right\}
\end{equation}

\noindent Because the gauge constraint contains a free Lorentz index, as
does the gauge transformation variable $\epsilon_{\beta}$, the ghost field
will carry a Lorentz label, i.e., they will be fermionic vector
fields.
After a
bit of work the ghost Lagrangian is found to be

\begin{equation}
{\cal L}_{gh} = \sqrt{ \bar{g}} \eta^{* \mu} \left[ D_{\lambda} D^{\lambda}
\bar{g}_{\mu \nu} - R_{\mu \nu} \right] \eta^{\nu}
\end{equation}

\noindent The full quantum action is then of the form

\begin{eqnarray}
S = \int d^4s \sqrt{\bar{g}} \left\{ {2 \over \kappa^2} \bar{R} - {1 \over 2}
h_{\alpha \beta} D^{\alpha \beta, \gamma \delta} h_{\gamma \delta} \right.
\nonumber \\
+ \left. \eta^{* \mu} \left\{ D_{\lambda} D^{\lambda} \bar{g}_{\mu
\nu} - \bar{R}_{\mu \nu} \right\} \eta^{\nu} + {\cal O}(h^3) \right\}
\end{eqnarray}

\noindent with $D^{\alpha \beta, \gamma \delta}$ an invertible differential
operator  formed using Eq.31 and Eq.36.  This can then be used to define the
propagator  and the Feynman rules in a straightforward fashion.

Despite the conceptual simplicity, explicit formulas in gravity have a
notational complexity due to the proliferations of Lorentz indices.  Around
flat space, the momentum space propagator is relatively simple in this gauge

\begin{eqnarray}
i D_{\mu \nu \alpha \beta} & = & {i \over q^2 + i \epsilon} P_{\mu \nu,
\alpha \beta} \nonumber \\
P_{\mu \nu, \alpha \beta} & \equiv & {1 \over 2} \left[ \eta_{\mu \alpha}
\eta_{\nu \beta} + \eta_{\mu \beta} \eta_{\nu \alpha} - \eta_{\mu \nu}
\eta_{\alpha \beta} \right]
\end{eqnarray}

\noindent The coupling to matter of the one graviton and two graviton
vertices are
respectively

\begin{eqnarray}
\tau_{\mu \nu} & = & -{i\kappa \over 2} \left( p_{\mu}
p^{\prime}_{\nu} +
p^{\prime}_{\mu} p_{\nu} - g_{\mu \nu} [p \cdot p^{\prime} - m^2]
\right)
\nonumber \\
\tau_{\eta \lambda, \rho \sigma} & = & {i \kappa^2 \over 2} \left\{
I_{\eta \lambda, \alpha \delta} I^\delta _{~\beta, \rho \sigma} \left(
p^{\alpha} p^{\prime \beta} + p^{\prime \alpha} p^{\beta} \right) \right.
\nonumber \\
& & - {1 \over 2} \left( \eta_{\eta \lambda} I_{\rho \sigma,
\alpha \beta} +  \eta_{\rho \sigma} I_{\eta \lambda, \alpha \beta} \right)
p^{\prime \alpha}  p^{\beta}  \nonumber \\
& & \left. - {1 \over 2} \left( I_{\eta \lambda, \rho \sigma} - {1 \over 2}
\eta_{\eta \lambda} \eta_{\rho \sigma} \right) [p \cdot p^{\prime} - m^2 ]
\right\}
\end{eqnarray}

\noindent with

\begin{equation}
I_{\mu \nu, \alpha \beta} \equiv {1 \over 2} [ \eta_{\mu \alpha} \eta_{\nu
\beta} + \eta_{\mu \beta} \eta_{\nu \alpha} ]
\end{equation}

\noindent The energy momentum tensor for gravitons leads to the interaction
of gravitons with an external field which, with the harmonic gauge
fixing, is of the form

\begin{eqnarray}
\tau^{\mu \nu}_{\alpha \beta, \gamma \delta} & = & { i\kappa \over 2} \left(
P_{\alpha \beta, \gamma \delta} \left[ k^{\mu} k^{\nu} + (k - q)^{\mu} (k-
q)^{\nu} + q^{\mu} q^{\nu} - {3 \over 2} \eta^{\mu \nu} q^2 \right]
\right. \nonumber \\
& &+ 2q_{\lambda} q_{\sigma} \left[ I^{\lambda \sigma,}_{~~~\alpha
\beta}
I^{\mu \nu,} _{~~~\gamma  \delta} + I^{\lambda \sigma,} _{~~~\gamma
\delta}
I^{\mu \nu,} _{~~~\alpha  \beta} - I^{\lambda \mu,} _{~~~\alpha \beta}
I^{\sigma \nu,} _{~~~\gamma \delta}  - I^{\sigma \nu,} _{~~~\alpha \beta}
I^{\lambda \mu,} _{~~~\gamma \delta}  \right] \nonumber \\
& &+ \left[ q_{\lambda} q^{\mu} \left( \eta_{\alpha \beta} I^{\mu \nu,}
_{~~~\gamma \delta} + \eta_{\gamma \delta} I^{\lambda \nu,} _{~~\alpha
\beta}  \right) + q_{\lambda} q^{\nu} \left( \eta_{\alpha \beta} I^{\lambda
\mu,}_{~~~\gamma \delta} + \eta_{\gamma \delta} I^{\lambda \mu,}
_{~~~\alpha \beta}  \right) \right. \nonumber \\
& & \left. -q^2 \left( \eta_{\alpha \beta} I^{\mu \nu,} _{~~~\gamma \delta}
+  \eta_{\gamma \delta} I^{\mu \nu,} _{~~~\alpha \beta} \right) - \eta^{\mu
\nu}  q^{\lambda} q^{\sigma} \left( \eta_{\alpha \beta} I_{\gamma \delta,
\lambda \sigma} + \eta_{\gamma \delta} I_{\alpha \beta, \lambda \sigma}
\right) \right] \nonumber \\
& & + \left[ 2q^{\lambda} \left( I^{\sigma \nu,} _{~~~\alpha \beta}
I_{\gamma  \delta, \lambda \sigma} (k - q)^{\mu} + I^{\sigma \mu,}
_{~`~\alpha \beta}  I_{\gamma \delta, \lambda \sigma} (k-q)^{\nu} \right.
\right. \nonumber \\
& & \left. -I^{\sigma \nu,} _{~~~\gamma \delta} I_{\alpha
\beta, \lambda  \sigma} k^{\mu} - I^{\sigma \mu,} _{~~~\alpha \beta,
\lambda
\sigma}  k^{\nu} \right) \nonumber \\
& & \left. +q^2 \left( I^{\sigma \mu,} _{~~~\alpha \beta} I_{\gamma \delta,
\sigma} ^{~~~~\nu} + I_{\alpha \beta, \sigma} ^{~~~~\nu} I^{\sigma
\nu,}
_{~~\alpha  \delta} \right) + \eta^{\mu \nu} q^{\lambda} q_{sigma} \left(
I_{\alpha  \beta, \lambda \rho} I^{\rho \sigma,} _{~~\gamma \delta} +
I_{\gamma \delta, \lambda \rho} I^{\rho \sigma,} _{~~~\alpha \beta} \right)
\right] \nonumber \\
& & + \left\{ \left( k^2 + (k-q)^2 \right)
\left(I^{\sigma \mu,} _{~~~\alpha  \beta} I_{\gamma \delta, \sigma}
^{~~~~\nu} + I^{\sigma \nu,} _{~~~\alpha \beta}  I_{\gamma \delta,
\sigma}
^{~~~~\mu} - {1 \over 2} \eta^{\mu \nu} P_{\alpha  \beta, \gamma \delta}
\right) \right. \nonumber \\
& & \left. \left. - \left( k^2 \eta_{\gamma
\delta} I^{\mu \nu,} _{~~~\alpha  \beta} + (k-q)^2 \eta_{\alpha \beta}
I^{\mu
\nu,} _{~~\gamma \delta} \right)  \right\} \right)
 \end{eqnarray}

As a simple example of the use of these Feynman rules,
consider the interaction of two
heavy masses.  This is written as

\begin{equation}
M = {1 \over 2} \tau_{\mu \nu} (q) D^{\mu \nu, \alpha \beta} (q)
\tau_{\alpha \beta} (q)
\end{equation}

\noindent Taking the nonrelativistic limit $p_{\nu} \sim (m, \vec{0})$,
and accounting for the normalization of the states leads us to

\begin{equation}
{1 \over 2m_1} {1 \over 2m_2} M = 4 \pi G {m_1 m_2 \over q^2}
\end{equation}

\noindent which leads to the usual potential energy function.

\begin{equation}
V(r) = - G{m_1 m_2 \over r}
\end{equation}

\noindent Of course, this is a classical result which did not require us to go
through the quantization procedure.  True quantum effects will be discussed
later.

\section{Quantum Effective Field Theory-Overview}

While the quantization of gravity proceeds in an almost identical fashion to
that of Yang Mills theory, it is what happens next which has been
troublesome for gravity.  Because of the dimensionful
coupling $\kappa$ and the
nonlinear interactions to all orders in $h_{\mu \nu}$, gravity does not
belong to the class of renormalizable field theories.  As we will see, loop
diagrams generate divergences which cannot be absorbed in only a
renormalization of G, but require increasing numbers of renormalized
parameters with increasing numbers of loops.  However this pattern is
typical of effective field theories and is not an obstacle to making quantum
predictions.

The procedure for carrying out this program involves the following steps.
The action is organized in full generality in an energy expansion.  The
vertices and propagators of the theory start with terms from the lowest order
Lagrangian, and higher order Lagrangians are treated as perturbations.  The
quantum corrections are calculated and since the ultraviolet divergences
respect the symmetry of the theory and are local, they can be absorbed into
the parameters of the action.  These renormalized parameters must be
determined from experiment.  The remaining relations between amplitudes
are the predictions of the theory.  In the following sections I describe in
more detail the steps sketched above.

\section{The Energy Expansion}

We have already partially discussed this in previous sections.  In addition to
the gravitational Lagrangian (with $\Lambda = 0$)

\begin{equation}
{\cal L}_g = \sqrt{g} \left\{ {2 \over \kappa^2} R + c_1 R^2 +
c_2 R_{\mu \nu} R^{\mu
\nu} + {\cal O} (R^3) \right\}
\end{equation}

\noindent the matter Lagrangian must also be written with general
couplings in increasing powers of the curvature

\begin{eqnarray}
{\cal L}_m & = & \sqrt{g} \left\{ {1 \over 2} (g^{\mu \nu} \partial_{\mu}
\phi \partial_{\nu} \phi - m^2 \phi^2) \right. \nonumber \\
& & + \left. d_1 R^{\mu \nu} \partial_{\mu} \phi \partial_{\nu} \phi + R
(d_2 g^{\mu \nu} \partial_{\mu} \phi \partial_{\nu} \phi + d_3 m^2 \phi^2
) + \ldots . \right\}
\end{eqnarray}

\noindent The couplings $c_i$ are dimensionless and in matrix elements the
expansion will be of the form $1 + Gq^2 c_i$.  The coefficients $d_i$ are a
bit more subtle.  They have dimension $d_i \sim 1/(mass)^2$.  In a theory
where gravity is the only low energy interaction, a point particle would be
expected to have $d_i$ of order $G$.  However for interacting theories or
composite particles the coefficients can be much larger.  In matrix elements
of the energy momentum tensor, $d_i$ play the role analogous to the charge
radius.  In QED, the photonic radiative corrections to the
energy-momentum charge radius of a charged particle
will generate $d_i$ of order $\alpha / m^2$.  In the case of
bound states, a composite particle will have
$d_i$ of order the physical spatial extent of the particle $d_i \sim <r^2>$.

If we just had gravity plus a single type of matter field, we could use the
equations of motion to eliminate some terms in these effective Lagrangian,
as the equations of motion relate the curvatures to the matter field.  When
treated as an effective field theory, it is fair to use the
lowest order equations
of motion to simplify the next order Lagrangian.  However, in practice we
have several types of possible matter fields, as well as interactions among
these fields, so that the equations of motion would vary according to which
fields were included.  I have therefore not eliminated any terms by the
equations of motion.

\section{Renormalization}

The one loop divergences of gravity have been studied in two slightly
different methods.  One involves direct calculation of the Feynman diagrams
with a particular choice of gauge and definition of the quantum gravitational
field [16].  The background field method,
with a slightly different gauge constraint, allows one to calculate in
a single step
the divergences in graphs with arbitrary numbers of external lines and
also produces a result which is explicitly generally covariant[4].
In the latter technique one expands about a
background spacetime $\bar{g}_{\mu \nu}$, fixes the gauge as we
described above and collects all the terms quadratic in the quantum field
$h_{\mu \nu}$ and the ghost fields.  For the graviton field we have

\begin{eqnarray}
Z \left[ \bar{g} \right] & = & \int \left[ dh_{\mu \nu} \right] exp \left\{ i
\int d^4 x \sqrt{\bar{g}}  \left\{ {2 \over \kappa^2} \bar{R} + h_{\mu \nu}
D^{\mu \nu \alpha \beta}  h_{\alpha \beta} \right\} \right. \nonumber \\
& = & det D^{\mu \nu \alpha \beta} \nonumber \\
& = & exp Tr ln (D^{\mu \nu \alpha \beta})
\end{eqnarray}

\noindent where $D^{\mu \nu \alpha \beta}$ is a differential operator made
up of derivatives as well as factors of the background curvature.  The short
distance divergences of this object can be calculated by standard techniques
once a regularization scheme is chosen.  Dimensional regularization is the
preferred scheme because it does not interfere with the invariances of
general relativity.  First calculated in this scheme
by 't Hooft and Veltman[4], the divergent
term at one-loop due to graviton and ghost loops is described by a
Lagrangian

\begin{equation}
{\cal L}^{(div)}_{1 loop} = {1 \over 8 \pi^2 \epsilon} \left\{ {1 \over 120}
\bar{R}^2 + {7 \over 20} \bar{R}_{\mu \nu} \bar{R}^{\mu \nu} \right\}
\end{equation}

\noindent with $\epsilon = 4 - d$.  Matter fields of different spins will also
provide additional contributions with different linear combinations of $R^2$
and $R^{\mu \nu} R_{\mu \nu}$ at one loop.

The fact that the divergences is not proportional to the original Einstein
action is an indication that the theory is of the non-renormalizable type.
Despite the name, however, it is easy to renormalize  the theory at any given
order.  At one loop we identify renormalized parameters

\begin{eqnarray}
c^{(r)}_1 & = & c_1 + {1 \over 960 \pi^2 \epsilon} \nonumber \\
c^{(r)}_2 & = & c_2 + {7 \over 160 \pi^2 \epsilon}
\end{eqnarray}

\noindent which will absorb the divergence due to graviton loops.  Alternate
but equivalent expressions would be used in the presence of matter loops.

A few comments on this result are useful.  One often hears that pure gravity
is one loop finite.  This is because the lowest order equation of motion for
pure gravity is $R_{\mu \nu} = 0$ so that the ${\cal O} (R^2)$ terms in the
Lagrangian vanish for all solutions to the Einstein equation.  However in the
presence of matter (even classical matter) this is no longer true and the
graviton loops yield divergent effects which must be renormalized as
described above.  At two loops, there is a divergence in pure
gravity which remains even after the equations of motion have been used [17].

\begin{equation}
{\cal L}^{(div)}_{2 loop} = {209 \kappa^2 \over 2880 (16 \pi^2 )} {1 \over
\epsilon} \bar{R}^{\alpha \beta}_{~~\gamma \delta} \bar{R}^{\gamma
\delta}_{~~\eta \sigma} \bar{R}^{\eta \sigma}_{~~\alpha \beta}
\end{equation}

\noindent For our purposes, this latter result also serves to illustrate the
nature of the loop expansion.  Higher order loops invariably involve more
powers of $\kappa$ which by dimensional analysis implies more powers of the
curvature or of derivatives in the corresponding Lagrangian (i.e., one loop
implies $R^2$ terms, 2 loops imply $R^3$ etc.).  The two loop divergence
would be renormalized by absorbing the effect into a renormalized value of
a coupling constant in the ${\cal O}(R^3)$ Lagrangian.

\section{Quantum Predictions in An Effective Theory}

At this stage it is important to be clear about the nature of the quantum
predictions in an effective theory.  The divergences described in the last
sections come out of loop diagrams, but they are {\em not} predictions of
the effective theory.  They are due to the high energy portions of the loop
integration, and we do not even pretend that this portion is reliable.  We
expect the real divergences (if any) to be different.  However the
divergences do not in any case enter into any physical consequences, as
they absorbed into the renormalized parameters.  The couplings which
appear in the effective Lagrangian are also not predictions of the effective
theory.  They parameterize our ignorance and must emerge from an ultimate
high energy theory or be measured experimentally.  However there are
quantum effects which are due to low energy portion of the theory, and
which the effective theory can predict.  These come because the effective
theory is using the correct degrees of freedom and the right vertices at low
energy.  It is these low energy effects which are the quantum predictions of
the effective field theory.

It may at first seem difficult to identify which components of a calculation
correspond to low energy, but in practice it is straightforward.  The
effective field theory calculational technique automatically separates the low
energy observables.  The local effective Lagrangian will generate
contributions to some set of processes, which will be parameterized by a set
of coefficients.  If, in the calculation of the loop
corrections, one encounters
contributions which have the same form as those from the local Lagrangian,
these cannot be distinguished from high energy effects.   In the comparison
of different reactions, such effects play no role, since we do not know
ahead of time the value of the coefficients in ${\cal L}$.  We must
measure these constants or form linear combinations of observables
which are independent of them. Only loop
contributions which have a different structure from the local
Lagrangian can make a difference in the
predictions of reactions.  Since the effective Lagrangian accounts for the
most general high energy effects, anything with a different structure must
come from low energy.

A particular class of low energy corrections stand out as the most important.
These are the nonlocal effects.  In momentum space the nonlocality is
manifest by a nonanalytic behavior.  Nonanalytic terms are clearly distinct
from the effects of the local Lagrangian, which always give results which
involves powers of the momentum.

Let us illustrate the nature of the quantum corrections by considering the
interactions of two loop heavy masses.  In coordinate space we can consider
possible power modifications at order G of the interactions of the form

\begin{equation}
V(r)=-{Gm_1 m_2 \over r} \left( 1 + a {Gm \over rc^2} + b {G \hbar \over
r^2 c^3} + \ldots \right)
\end{equation}

\noindent The form of these corrections is fixed strictly by dimensional
analysis.  The first, ${Gm \over rc^2}$, is the classical expansion parameter
for the nonlinear effects in classical general relativity.  In contrast, the
second is the unique form linear in $G \hbar$ and is the quantum expansion
parameter.  In momentum space, obtained by the Fourier transform of the
potential, one has the corresponding expansion (up to constants of order 1)

\begin{equation}
V(q) \sim {Gm_1 m_2 \over q^2} \left( 1 + a G q^2 \sqrt{{m^2 \over -
q^2}} + b G q^2 ln(-q^2) + \ldots \right)
\end{equation}

\noindent The nonanalytic term of the form $G q^2 \sqrt{{m^2 \over -
q^2}}$ corresponds to the classical expansion parameter, while the
nonanalytic form $Gq^2 ln(-q^2)$ corresponds to the quantum expansion
parameter.  If we now turn to a calculation in an
effective field theory (to be
described in the next section) we directly find the desired terms, with an
analytic contribution also

\begin{equation}
V(q) \sim {Gm_1 m_2 \over q^2} \left( 1 + a G q^2 \sqrt{{m^2 \over -
q^2}} + b G q^2 ln -q^2 + c G q^2 + \ldots \right)
\end{equation}

\noindent The analytic term $Gq^2$ receives contributions from the local
effective Lagrangian, and also from the quantum loops.  It is therefore not a
quantum prediction of the low energy effective theory.  On the other hand
the one loop calculation shows that the nonanalytic terms are finite and
independent of the coefficients in the effective Lagrangian (aside from G).
These are then predictions of the low energy theory.  Note that the analytic
term Fourier transforms to a delta function $\delta V \sim G^2 m_1 m_2
\delta^3 (r)$, which at finite $r$ is smaller than any power  correction.
Thus the leading power corrections to the gravitational potential are reliably
predicted by the effective theory, including the quantum correction!

\section{ Quantum corrections to the Gravitational Potential}

The only complete example of this program is the calculation
of the effect of quantum physics on the  gravitational interaction
of two heavy masses. The power-law corrections to the
usual  ${1 \over r}$ potential are calculable. While there is
more than one way to define what one means by the potential
when one is working beyond leading order[5,18], the calculation of
the quantum corrections to that object are well defined. I
will describe the specific one-particle-reducible potential[5]
defined by including the vertex and propagator modifications
of one graviton exchange.

The vertex corrections for a scalar particle have the most general form

\begin{eqnarray}
V_{\mu \nu} \equiv <p' \mid T_{\mu \nu} \mid p > & = & F_{1} (q^2)
\left[
p_{\mu} p'_{\nu} +
p'_{\mu} p_{\nu} + q^2 {\eta_{\mu \nu} \over 2} \right] \nonumber \\
& & + F_2 (q^2) \left[ q_{\mu} q_{\nu} - g_{\mu \nu} q^2 \right]
\end{eqnarray}

\noindent with $F_1 (0) = 1$. Including the contributions of higher
order effective Lagrangians  as well as graviton loops one
will obtain corrections of the form

\begin{eqnarray}
F_1 (q^2) & = & 1 + d_1 q^2 + \kappa^2 q^2 \left( \ell_1 + \ell_2 ln {(-
q^2) \over
\mu^2} +
\ell_3 \sqrt{{m^2 \over -q^2}} \right) + \ldots \nonumber \\
F_2 (q^2) & = & -4 (d_2 + d_3) m^2 +  \kappa^2 m^2 \left( \ell_4 + \ell_5
ln {(-
q^2) \over
\mu^2} +
\ell_6 \sqrt{{m^2 \over -q^2}} \right) + \ldots
\end{eqnarray}

\noindent where the $d_i$ are defined in Eq. 47 and
the  $\ell_i (i = 1, 2 \ldots 6)$ are
dimensionless numbers from the loop diagrams. This general
structure can be gotten from dimensional analysis. Note that
the $d_i$ can be interpreted as the charge radii for the
energy-momentum tensor. For the propagator correction,
allow me to drop Lorentz indices here in order to see the
physics rather than the indices. In the vacuum polarization
there is no mass, hence no correction of the form
$\sqrt{{ m^2 \over -q^2 }}$.
The propagator plus vacuum polarization is of the form

\begin{equation}
{1 \over q^2} + {1 \over q^2} \pi (q^2) {1 \over q^2} + \ldots = \left\{ {1
\over
q^2} + \kappa^2
\left[
c_1 + c_2 + \ell_7 + \ell_8 ln (-q^2) \right] \right\}
\end{equation}

 With the definition of
the one-particle-reducible interaction, we form

\begin{eqnarray}
& - & {\kappa^2 \over 4} {1 \over 2m_1} V^{(1)}_{\mu \nu} (q) \left[ i
D^{\mu
\nu, \alpha
\beta} (q) + i
D^{\mu \nu, \rho \sigma} i \Pi_{\rho \sigma, \eta \lambda} i D^{\eta
\lambda,
\alpha \beta} \right]
V_{\alpha \beta} (q) {1 \over 2m_2} \nonumber \\
& \approx & 4 \pi G m_1 m_2 \left[ {i \over {\bf q}^2} - {i \kappa^2 }
 \left[ a  ln ({\bf q}^2) + {b (m_1 + m_2) \over  \sqrt{{\bf q}^2}} \right]
+ const
\right]
\end{eqnarray}
\noindent where $a$ is a linear combination of $\ell_2,\ell_5, \ell_8$
and $b$ is a combination of $\ell_3,\ell_6$. The
factors of ${ 1 \over 2m }$ account
for the normalization used in our states.
When this is turned into a coordinate space potential the constant
terms yield zero at any finite radius, since

\begin{equation}
\int {d^3 q \over (2 \pi)^3} e^{-i \vec{q} \cdot \vec{r}} = \delta^3 (x)
\end{equation}

\noindent while the non analytic terms however lead to {\em power law}
behavior since

\begin{eqnarray}
\int {d^3q \over (2 \pi)^3} e^{-i \vec{q} \cdot \vec{r}} {1 \over q} = {1
\over 2
\pi^2 r^2}
\nonumber \\
\int {d^3q \over (2 \pi)^3} e^{-i \vec{q} \cdot \vec{r}} ln \vec{q}^2 = {-1
\over 2
\pi^2 r^3}
\end{eqnarray}
\noindent From our
discussion in the previous section, we know that
constant terms in $V(r)$ correspond to a delta function
potential and can be dropped at any finite $r$. The factors
of $\sqrt{{m^2 \over -q^2}}$ lead to the classical power-law
corrections and the $ln(-q^2)$ terms lead to the quantum
power-law corrections, as in Eq. 52. Therefore we need to calculate the
constants $\ell_2, \ell_5$ and $\ell_8$ and $\ell_3 , \ell_6$.

The calculation is a bit tedious because of all the Lorentz
indices, but one finds for the non-analytic terms in the vertex[3,5]

\begin{eqnarray}
F_1(q^2) &=& 1 + {\kappa^2 \over 32 \pi^2} q^2 \left[ - {3 \over 4} ln (-
q^2) +
{1 \over 16}
{\pi^2
m \over \sqrt{-q^2}} \right] \nonumber \\
F_2(q^2) &=& {\kappa^2 m^2 \over 32 \pi^2} \left[ - {4 \over 3} ln(-q^2)
+ {7
\over 8} {\pi^2 m
\over \sqrt{-q^2}} \right]
\end{eqnarray}

The non-analytic terms in the vacuum polarization can
be obtained from the work of 't Hooft and Veltman[4] by
noting that, is a massless theory, the $ln(-q^2)$ terms
are always  related to the coefficient of  ${1 \over d-4}$
in dimensional regularization. The appropriate combination is

\begin{eqnarray}
P_{\mu \nu, \alpha \beta} \Pi^{\alpha \beta, \gamma \delta} P_{\gamma
\delta, \rho
\sigma} =
{\kappa^2 q^4 \over 32 \pi^2} \left[ {21 \over 120} \left( \eta_{\mu \rho}
\eta_{\nu
\sigma} +
\eta_{\nu
\rho} \eta_{\mu \sigma} \right) + {1 \over 120} \eta_{\mu \nu}
\eta_{\rho \sigma} \right] \nonumber \\
\left[ -ln  (-q^2) \right] + \ldots
\end{eqnarray}

This leads to

\begin{eqnarray}
& - & {\kappa^2 \over 4} {1 \over 2m_1} V^{(1)}_{\mu \nu} (q) \left[ i
D^{\mu
\nu, \alpha
\beta} (q) + i
D^{\mu \nu, \rho \sigma} i \Pi_{\rho \sigma, \eta \lambda} i D^{\eta
\lambda,
\alpha \beta} \right]
V_{\alpha \beta} (q) {1 \over 2m_2} \nonumber \\
& \approx & 4 \pi G m_1 m_2 \left[ {i \over {\bf q}^2} - {i \kappa^2 \over
32
\pi^2} \left[ - {127
\over
60}  ln ({\bf q}^2) + {\pi^2 (m_1 + m_2) \over 2 \sqrt{{\bf q}^2}} \right]
+ const
\right]
\end{eqnarray}

and the corresponding Fourier transform

\begin{equation}
V(r) = - {Gm_1 m_2 \over r} \left[ 1 - {G(m_1 + m_2) \over r c^2} - {127
\over
30 \pi^2} {G
\hbar \over r^2 c^3} \right]
\end{equation}

\noindent Photons and massless neutrinos also modify the vacuum
polarization at large distance, generating $ln(-q^2)$
corrections[16]. If we include these, with $N_\nu$ massless
neutrinos,  the quantum piece becomes

\begin{equation}
 - {135 +2 N_\nu
\over
30 \pi^2} {G
\hbar \over r^2 c^3}
\end{equation}

  When we perform a perturbative expansion, we want the
corrections to be small and well behaved. In this regard,
gravity is the best perturbative theory in Nature! The
quantum correction ${G {\hbar} \over r^2c^3}$ is
about $10^{-38}$ at $r = 1$ fm, and is unmeasurably tiny
on macroscopic scales. This of course is a success; we want
quantum gravity to have a well behaved classical limit.
However it indicates that this is unlikely to generate an
active phenomenology. Overall though, the number is not
important; the methodology is. Quantum predictions can be
made in general relativity.

\section{Issues in the Structure of Gravitational Effective Field Theory}

In this section I would like to describe two ways in
which the gravitational effective field theory superficially
appears different from standard expectations in other
effective field theories.

\noindent i) The extreme low energy limit.

      The considerations above concern what could be
called ``ordinary'' distance scales, i.e. fermis to kilometers,
and regions where the curvature is small.  Certainly as
one goes to high energies, the methodology is no longer applicable.
However, gravity may also behave oddly in the limit of {\it {extreme}}
low energy, i.e. as the wavelength probed becomes comparable
to the size of the universe or the distance to the nearest
black hole. For example, there are singularity theorems[19] which
state that for most reasonable matter distributions,
spacetime evolved by Einstein's equations has a singularity
in the future or past. The singularity itself is not
necessarily a problem; most likely the singularity is
smoothed out in the full high energy theory. However, it
is unusual that the low energy theory evolves into a state where
it is no longer valid. The G.E.F.T. would not work in the
neighborhood of a singularity. Therefore we could have a
reasonable calculation which works for some ordinary wavelengths
but cannot be applied for
$\lambda \to \infty$ because of the presence of large curvature. The
existence of a horizon around black holes may also be a
problem as $\lambda \to \infty$. The horizon itself is not a
problem, in the sense that the curvature can be very small at
the horizon so that a freely falling observer would be able to
apply the effective theory locally. However, problems
associated with the horizon could possible arise as
$\lambda \to \infty$ since regions at spatial infinity
are inaccessible to processes inside the horizon. I do not
know if in fact there are real problems  with the
infinite wavelength limit, but at least this is a
limit where our past experience with effective theories
is not applicable. In many ways, it would be more interesting
to have a problem at the extreme low energy end rather
than the better known problems at high energy, since
we could not rely on new physics at high scales to resolve the issue.

ii) Naive power counting

 The second point involves a technical issue, i.e. the
correct counting of momenta in loops. In chiral theories,
Weinberg[20] proved an elegant power counting theorem which
shows that higher order loop diagrams always generate
higher orders in the energy expansion. Explicit calculations
in gravity have also followed this pattern. However, there does
not yet exist the equivalent of Weinberg's power counting
theorem. In fact an attempt to do naive power counting runs
into an obstacle in that it seems to give a dangerous behavior
for some diagrams. For example if one calculates the box diagram
where two heavy particles interact by the exchange of two
gravitons, both naive power counting and explicit calculation
give a correction of the form
\begin{equation}
{\kappa^2 m_1^2 m_2^2 \over q^2}(\kappa^2m^2)
\end{equation}

\noindent i.e. an apparent expansion in the dimensionless
variable $\kappa^2m^2$. However this would be a disaster
as $\kappa^2m^2$ can be large for classical objects and
because it does not allow a classical limit. When one puts in
factors of $\hbar$ and $c$ one
has ${\kappa^2 m^2c \over \hbar}$, i.e.
the $ \hbar$ is in the denominator. Direct calculation
shows that this bad behavior is canceled by the crossed
box diagram, and that the sum of the two fulfills standard
expectation. This is similar to the cancellation if infrared
divergences, which is known to occur in gravity in the same
way as in QED[21]. With some work, some of the systematics of this
cancellation in other diagrams can be worked out, and T. Torma and
I will soon have a paper describing these issues[22].  We
have not found any breakdown of the standard expectation
of the nature of the quantum expansion; but the field
still lacks a general proof comparable to Weinberg's.

\section{Future Directions}

The gravitational effective field theory provides a new
technology for quantum gravity. Most discussions of the
topic of gravity and quantum mechanics do not keep track
of the high vs low scales. This ``effective'' way of thinking
can be very useful in deciding which aspects are trustworthy
and which are speculative.

Within the effective theory, there are several possible
directions for future work. It may be possible to compare
the quantum predictions with the results of computer
simulations of lattice gravity. It is too early to entirely
give up on  all hopes for real phenomenology. Perhaps quantum
effects can build up and be visible as deviations from
various null effects of the classical  theory[23]. Perhaps these
ideas may be useful in describing the very early universe.
There are also potential theoretical applications, such as
the effects of anomalies or the quantum influence on the
development of singularities. Finally there exist in the
literature various suggestions for unusual gravitational
effects such as phase transitions, running $G$ at low energy,
solutions to the dark matter problem etc. Effective field
theory should be able to support or refute these suggestions.
At the least, we will put our standard expectations on a
stronger footing.

Nature has apparently given us the fields and interactions of the
electroweak gauge theory, quantum chromodynamics and general
relativity at present energies. All are treatable at those energies
by the techniques
of quantum mechanics. It remains a formidable challange to
construct the ultimate high energy structure of Nature. We
expect it to be quite different from what we presently have.
Because physics is an experimental science, it is possible that we
will not be able to truly know the ultimate theory, but impressive
attempts are underway. However we do have a right to expect that
our theories are consistent at the energies that we use them.
In this regard, the effective field theory framework is the
appropriate description of quantum gravity at ordinary energies.

\begin{center}
{\bf Acknowledgement}
\end{center}

I would like to thank the organizers and participants
of this summer school for their hospitality and scientific discussions.
This work has been partially supported by the U. S. National  Science
Foundation.

\end{document}